\def\cm{{\rm\thinspace cm}}

\def\erg{{\rm\thinspace erg}}

\def\K{{\rm\thinspace K}}
\def\keV{{\rm\thinspace keV}}
\def\km{{\rm\thinspace km}}

\def\Msun{\hbox{$\rm\thinspace M_{\odot}$}}
\def\pc{{\rm\thinspace pc}}

\def\s{{\rm\thinspace s}}
\def\yr{{\rm\thinspace yr}}

\def\pcmcu{\hbox{$\cm^{-3}\,$}}

\def\ergpcmsqps{\hbox{$\erg\cm^{-2}\s^{-1}\,$}}

\def\ergps{\hbox{$\erg\s^{-1}\,$}}

\def\kmps{\hbox{$\km\s^{-1}\,$}}

\def\Msunpyr{\hbox{$\Msun\yr^{-1}\,$}}
\def\pcm{\hbox{$\cm^{-3}\,$}}

\documentclass[usegraphicx]{mn2e}

\usepackage{fixltx2e}
\usepackage{mathptmx}

\include{defn}
\begin{document}

\title[Power to the filaments]{The energy source of the filaments
  around the giant galaxy NGC\,1275} \author[Fabian et al]
{\parbox[]{6.5in}{{A.C. Fabian$^1\thanks{E-mail: acf@ast.cam.ac.uk}$,
      J.S. Sanders$^1$,
      R.J.R. Williams$^2$, A. Lazarian$^3$,  G.J. Ferland$^4$  and R.M. Johnstone$^1$}\\
    \footnotesize
    $^1$ Institute of Astronomy, Madingley Road, Cambridge CB3 0HA\\ 
$^2$ AWE plc, Aldermaston, Reading RG7 4PR\\
$^3$ Astronomy Department, University of Wisconsin, Madison, WI 53706, USA\\
$^4$ Department of Physics, University of Kentucky, Lexington, KY
40506, USA\\
\\
Contains Material (c) Crown Copyright MoD/2011\\
  }}
\maketitle
  
\begin{abstract}
  The brightest galaxy in the nearby Perseus cluster, NGC\,1275, is
  surrounded by a network of filaments. These were first observed
  through their H$\alpha$ emission but are now known to have a large
  molecular component with a total mass approaching $10^{11}\Msun$ of
  gas. The filaments are embedded in hot intracluster gas and stretch
  over 80~kpc. They have an unusual low excitation spectrum which is
  well modelled by collisonal heating and ionization by secondary
  electrons. Here we note that the surface radiative flux from the
  outer filaments is close to the energy flux impacting on them from
  particles in the hot gas. We propose that the secondary electrons
  within the cold filaments, which excite the observed submillimetre
  through UV emission, are due to the hot surrounding gas efficiently
  penetrating the cold gas through reconnection diffusion. Some of the
  soft X-ray emission seen from the filaments is then due to charge
  exchange, although this is insufficient to account for all the
  observed X-ray flux. The filaments are complex with multiphase
  gas. Interpenetration of hot and cold gas leads to the filaments
  growing in mass, at a rate of up to $100\Msunpyr$. The lack of soft
  X-ray cooling emission in cool core clusters is then due to the
  non-radiative cooling of hot gas on mixing with cold gas around and
  within the central galaxy.
\end{abstract}


\section{Introduction}

Networks of filaments are seen surrounding the central galaxy in some
cluster of galaxies (Baade \& Minkowski 1954; Lynds 1972; Johnstone,
Fabian \& Nulsen 1987; Heckman et al 1989; Crawford et al 1999). A
relatively nearby and spectacular example is the filamentary
nebulosity around NGC\,1275 at the centre of the Perseus cluster of
galaxies. This has been extensively studied with recent images of the
bright H$\alpha$ emission being published by Conselice et al (2001)
and Fabian et al (2008).  The origin and means of excitation of these
filaments remain a matter of debate.  They are unlikely to be
photoionized since the known sources of ionizing radiation are not
luminous or distributed enough.  The emission spectrum of the
filaments is also unlike that of any known Galactic nebula (although
it bears comparison with the Crab Nebula).  Strong molecular emission
is also present (Edge 2001, Edge et al 2002, Donahue et al 2000, Hatch
et al 2005) in this and similar brightest cluster galaxies. H$_2$
vibration temperatures of several thousand degreees are inferred
(Wilman et al 2002; Jaffe et al 2001).  CO emission is mapped across
the galaxy and even the outer filaments (Salome et al 2008), yielding
a total molecular mass for the filament system approaching
$10^{11}\Msun$.  The photoionization models which come closest to
reproducing the spectrum do so with high-energy photons (Crawford \&
Fabian 1992).

Johnstone et al (2007) used Spitzer and ground-based observations to
study characteristics of the H$_2$ emission seen in these filaments.
They found a correlation between level excitation energy and the
derived excitation temperature.  A follow-up study showed that the
collisional heating from ionizing particles could account for this
correlation, if a range of heating rates were present (Ferland et al
2008).  Agreement to within a factor of two over a wide range of line
ratios is good evidence that the model is viable.  The
optical/IR/radio emission-line spectrum most likely originates in gas
that is exposed to ionizing particles (Ferland et al 2009).

There are several possible sources of these ionizing particles
including true cosmic rays, relativistic particles produced {\em in
  situ}\/ by MHD phenomena, energetic photoelectrons produced by X-ray
photoionization, and penetration by particles originating in the
surrounding hot gas.  Here we argue that the latter is the most likely
energy source, due to the coincidence in the energy flux of hot
thermal particles onto the filament and the total atomic and molecular
emission from the filament. 

To explore and test the mechanism we concentrate on a Northern
filament in the Perseus cluster about 10~kpc north of the nucleus.  It
is clearly resolved in deep Chandra observations which show that it
has a width and surface brightness distribution very similar to that
of the underlying H$\alpha$ feature seen in the WIYN map by Conselice
et al (2001) or the Hubble Space Telescope images of Fabian et al
(2008); see Fig.~1 and Sanders \& Fabian (2007).  In Section 2 we
address the energetics and penetration of the hot and cold gases. In
Section 3 we present and discuss the Chandra X-ray spectrum of the
filament.  The X-ray spectrum may be consistent with that
expected from charge exchange as the highly ionized intracluster
medium recombines in contact with the much colder, and much more
neutral gas in the filament. The predicted X-ray flux, is however one
to two orders of magnitude too faint to explain the observations.  We
use a spectrum from the XMM-Newton Reflection Grating Spectrometer
(RGS) of the cluster core to show that cool X-ray line emission is
indeed present, although that spectrum is of a much larger scale than``s
any single filament.  We discuss turbulent diffusive reconnection as a
means to allow interpenetration to occur in Section 4.  Finally we
discuss wider implications of the model.

\section {The energetics and penetration of a cold filament by
  surrounding hot gas}

We begin by considering the cold filament to be a fixed surface
embedded in a much hotter medium, the intracluster gas, and assume
thermal pressure balance across the surface. Thermal conduction has
been considered several times for transporting thermal energy from the
hot to the the cold gas (B\"{o}hringer \& Fabian 1998; Nipoti \&
Binney 2005; Sparks et al 2009). The cold and hot faces do not
interact via classical thermal conduction, since this would lead to a
thick interface many kpc thick (e.g. B\"{o}hringer \& Fabian 1998)
which is not seen in X-ray images of the Perseus cluster (Sanders \&
Fabian 2007; see also Fig.~1).  Chandra observations of the hot gas
around the Northern filament of NGC\,1275 indicate an electron density
of $n_{\rm e}\sim 0.035\pcmcu$ and a temperature of about 4~keV right
up to the filaments, where the temperature drops to around 1~keV (see
Section 3).  The interaction between the hot and cold components is
therefore much more intimate.  Saturated conduction would then be
appropriate, with a heat flux given by Cowie \& McKee (1977) of $5\phi
p c_{\rm s} $, where $\phi\sim 1$ accounts for the uncertain physics,
$p$ is the gas pressure and $c_{\rm s}$ is the sound speed of the
gas. This is about $0.1 \ergpcmsqps$ for the above gas. Note however
that the estimate ignores magnetic fields. It is close however to
  the energy flux in the sonic limit. 

  Once in the cold gas phase, the (now) suprathermal particles rapidly
  lose energy by elastic collisions with free electrons, and by
  several inelastic processes including excitation, ionization, and
  dissociation of atoms and molecules (Spitzer \& Tomasko 1968;
  Dalgarno et al 1999).  Secondary ``knock-on'' electrons resulting
  from the initial collision of the energetic suprathermal particles
  cool by elastic collisions with cool electrons in (mixing) regions
  when the electron fraction is greater than a few percent cent.  They
  predominantly cool by inelastic collisions when the ionization
  fraction is lower.  Cloudy uses numerical fits to the Dalgarno et al
  results to follow the energy loss, excitation, and
  ionization/dissociation in detail. 

The colder gas has a range of temperatures from around $2\times
10^4\K$ and below with an observed H$\alpha$ flux of $\sim 5\times
10^{-15}\ergpcmsqps{\rm arcsec}^{-2}$ (Conselice et al 2001),
corresponding to an emitted flux of $7\times 10^{-4}\ergpcmsqps$.  The
total emitted surface flux, including Ly$\alpha$ in the UV to
molecular emission in the IR and submm is likely to be 20 times larger
or $\sim 10^{-2}\ergpcmsqps$.  This is interestingly close to that of
the impinging thermal particles provided that the excitation
efficiency $f$ with which they penetrate and excite the cold gas is of
order 0.1.

Magnetic fields must play an important role in maintaining the
integrity of the filaments and in supporting them against infall in
the cluster potential (Fabian et al 2008). This will also prevent
instant penetration of particles.  

If the media were laminar then the magnetic field would present a
major obstacle for diffusion perpendicular to the magnetic field. The
effective mean free path of hot particles perpendicular to the
magnetic field would be limited to the Larmor radius of hot particles
and, even in the case of the maximal rate of diffusion given by the
Bohm rate, it is much lower than for non-magnetized plasmas. However,
the natural state of astrophysical fluids is turbulent. The diffusion
of turbulent fluids perpendicular to the mean magnetic field is
typically many orders higher than the laminar estimates (see Lazarian
2006).

The sources of turbulence in intracluster medium are numerous, for
instance they may be related to the instabilities between hot and cold
gas (e.g. Begelman \& Fabian 1990; Knoll \& Brackbill 2002; Matsumoto
\& Hoshimo 2006). Without specifying the source of
turbulence, it is natural to assume that, on the size scale we are
dealing with, the high Reynolds number plasma is in a
turbulent state.


It is important to note that the detailed geometry of a filament is
complex.  The Northern filaments shown in Fig.~1 (right) is about 1~kpc
across and is made of threads which are resolved by HST at about 70~pc
diameter.  Even these cannot be filled by cold gas since their mean
density is $\sim 2\pcmcu$ (Fabian et al 2008) whereas the CO emitting
molecular gas (mostly H$_2$) is $\sim 10^5\pcmcu$), indicating a
volume filling factor $f_{\rm V}\sim 10^{-5}$. If we assume that the
cold gas occupies many strands of radius $r$ within a thread, which is
of radius $R\sim 35\pc$, then the area covering fraction, $f_{\rm
  A}=f_{\rm V} (R/r).$ Presumably $f_{\rm A}>1$ in order that the heat
flux into a thread can match the radiated flux, so $r<10^{15}\cm$.
  
It is beyond the scope of the present work to investigate the details
of the microscopic processes involved. Here we motivate reasons why
the penetration factor $f$ with which the hotter gas particles
penetrate the cold gas may be high.  If, as a result, the hot
particles reach the magnetic field lines which penetrate the cold gas
then this will allow them to generate enhanced emission by the
processes modelled by Ferland et al (2009).  The radiative efficiency
of the cold gas is so high (with cooling times very much less than
that of the hot gas) that we can largely ignore evaporation of the
cold gas (but see Section 3.2.2), which must be slowly accreting mass
from the hot phase.

\section{The X-ray spectrum and optical surface brightness of the
  Northern filament}

We now examine  the X-ray emission from the filaments, using the
Chandra spectrum of the Northern filament (shown in Fig.~2). The
region is $4.1 \times 24.3$ arcsec and background has been subtracted
using the neighbouring region shown in the Figure.  The
spectrum is well fit ($\chi^2=332/364$) by a two temperature model
with components at $0.66^{+0.07}_{-0.13}\keV$ and $1.46\pm0.26\keV$.
The abundance of the metals is $0.41^{+0.29}_{-0.15}$ (all
uncertainties are at the 90 per cent confidence level). The spectrum is
dominated by line emission: we still obtain an acceptable fit
($\chi^2=367/364$) if the abundance is raised to 10, indicating that the
evidence for any X-ray continuum in the filaments is weak. As shown in
Sanders \& Fabian (2007), the width of the filament is very similar in
the X-ray and visible bands.

The use of a two temperature model does not necessarily mean that all
the material present is exactly at the temperatures included.
It is perhaps more realistic to assume that there is a continuous
range of temperatures leading to emission lines from Si{\sc\,xii} to
Fe{\sc\,xvii}.  To check which lines are present, we have examined the
XMM RGS data from the core of the cluster.

\subsection{The XMM-Newton RGS spectrum of the Perseus cluster core}

The RGS data is from a slitless spectrometer with an aperture of 1 x 4
arcmin.  This is much too large to examine any particular region like
the Northern filament in detail, but the resultant spectrum again
demonstrates that the main features in the spectrum, apart from a
bremsstrahlung continuum and a significant contribution from the
jetted nucleus, are the expected emission lines.  These include
Fe{\sc\,xvii} which generally indicates emission from gas at $\sim
0.5\keV$.

We conclude from the Chandra spectrum of the filament that the
X-ray spectrum is dominated by emission lines, particularly from Fe-L
and oxygen.
  
\begin{figure}
  \centering
  \includegraphics[width=0.97\columnwidth,angle=0]{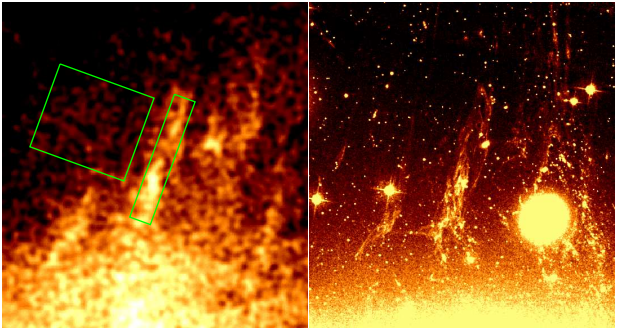}
  \caption{Chandra image of the Northern filament from which the
    spectrum is measured (right). The long
    box is 4.1 x 24.3 arcsec (1.5 x 9~kpc). The base of it is about
    24.4~kpc from the nucleus of NGC\,1275. }
\end{figure}

\begin{figure}
  \centering
  \includegraphics[width=0.7\columnwidth,angle=-90]{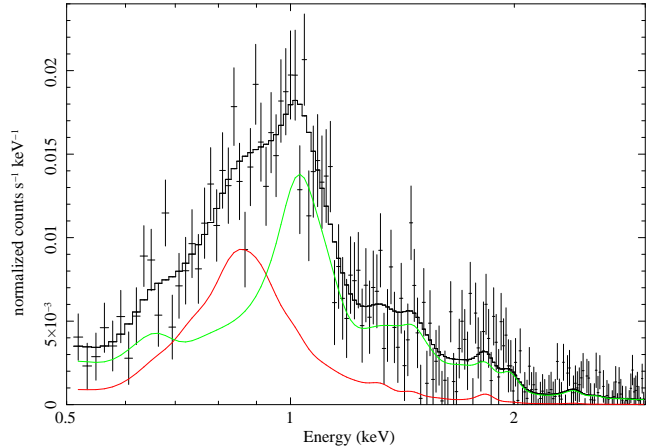}
  \caption{Top: Chandra spectrum of the filament region shown in
    Fig.1. Background has been subtracted. Bottom: Model of two
    temperature fit shown in Fig.2}
\end{figure}

\begin{figure}
  \centering
  \includegraphics[width=0.95\columnwidth,angle=0]{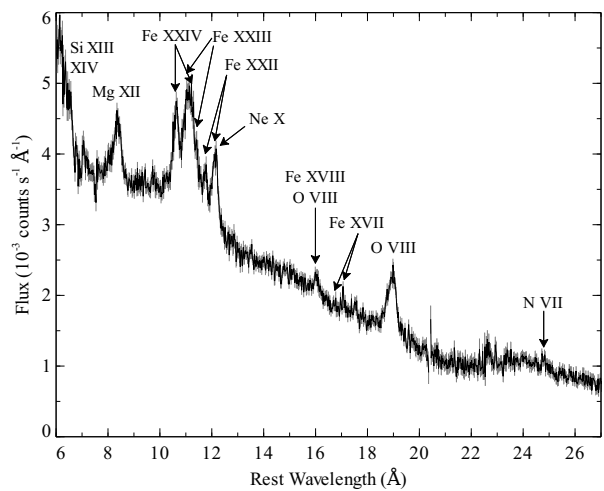}
  \caption{XMM-Newton RGS spectrum of the large region covering the
    cluster core.}
\end{figure}

\subsection{The energetics of filament X-ray emission}

\subsubsection{Thermal emission from hot gas}

In an earlier analysis involving a single temperature for the X-ray
cooler gas, Sanders \& Fabian (2007) deduced that the filament X-ray emission
can be produced by a sheath of 0.7~keV gas with depth $\sim 10\pc$.
This could be the consequence of the mixing of some of the hotter gas
with the cold filament gas, but there is no obvious reason for
obtaining that intermediate temperature. A mixing layer predicts
(Begelman \& Fabian 1990) gas an order of magnitude cooler at the
geometric mean of the outer, hotter and inner, cooler temperatures
(i.e. below $10^6\K$). The microphysics of such a layer is however
very complex and beyond detailed consideration here. Instead we
consider other processes which may be relevant.

\subsubsection{A simple approach to charge exchange emission}

We now examine what we expect to observe from hot ionized gas
penetrating and mixing with cold neutral gas.  The process of charge
exchange will dominate with electrons from the neutral gas
transferring into excited levels of the impinging ions.  The
subsequent cascade will result in X-ray line emission.  This process
is observed in other environments.  For example, it explains the X-ray
emission from comets seen in our Solar System.  Highly-charged ionic
species in the Solar Wind interact with the hydrogen halo surrounding
a comet to produce a rich and luminous X-ray spectrum (Dennerl et al 1997).
The process also causes emission from Jupiter and other planets
(e.g. Branduardi-Raymont et al 2007) and from the interaction at the
heliosphere and has even been suggested for intracluster filaments
(Lallement 2004).

To estimate the energetics in a simple manner for the situation of a
cold filament embedded in the hot intracluster medium, we assume that
each ion recombines successively, thereby releasing X-ray photons.
The main components are iron which leads to 8 L-shell lines in the
0.7-1.2~keV band and oxygen which lead to two.  Multiplying by the
relative abundances of iron and oxygen ($4.7\times 10^{-5}$ and
$8.5\times 10^{-4}$ relative to hydrogen by number), and assuming
their lines emerge around 1 keV on average gives us a fraction of the
energy flux of $5\times 10^{-4}$, or a flux of $5\times10^{-5}
\ergpcmsqps$, which is ten per cent of the H$\alpha$ flux of $7\times
10^{-4}\ergpcmsqps$. The observed soft Xray flux is similar to the
H$\alpha$ flux in the Northern filament (Sanders \& Fabian 2007). The X-ray
flux from simple charge exchange fails by a further factor of two to
three when the observed metallicity of the gas ($Z\sim 0.5$) and
precise line energies are considered.

Charge exchange predicts soft X-ray emission in the band where
emission is observed, but in its simplest form fails by a factor of
one to two orders of magnitude (but see discussion in Section 5).
The contribution
of charge exchange may eventually be discerned with high spectral
resolution spectroscopy, since there are characteristic spectral
differences from recombination (e.g. Brown et al 2009), including a
lack of a continuum.

It is possible that the X-ray emission is thermal and due to mixing of
the hot and cold phases resulting in gas at $7-15 $ million degrees in
the space between strands.  Additional gas at a warm temperature of
around $5\times 10^5\K$ is implied by the detection of OVI emission
(Bregman et al 2006) from an aperture covering the central 6.2 kpc of
NGC\,1275. The filament is therefore highly multiphase, with
components ranging from 50 to $10^7\K$.

\subsection{The optical line surface brightness of filaments}

\begin{figure}
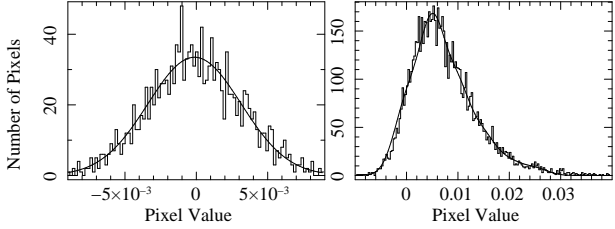

  \centering
  \includegraphics[width=0.35\columnwidth,angle=-90]{background.ps}
  \includegraphics[width=0.35\columnwidth,angle=-90]{composite_histogram.ps}
  \caption{ Distribution of pixel values (arbitrary units) in HST
    F625W optical image for a background region (left) and the
    Northern Filament (right). A model consisting of a 6 gaussians has
    been fitted on the right; the standard deviation of each is fixed
    to be the same as for the background and the pixel value of each
    steps in equal increments of $5\times 10^{-3}$ from 0. }
\end{figure}
We have investigated the surface brightness distribution in the
emission-line filaments using the HST ACS images in the F625W filter
published by Fabian et al (2008). These images are particularly
sensitive to emission from the H$\alpha$ and [NII] doublet. The
[SII]$\lambda\lambda$6717,6735 doublet also lies within the bandpass of
the filter but is rather fainter. The filter also transmits continuum
light from the galaxy.

We first subtracted a heavily smoothed copy of the F625W image from the
unsmoothed data in order to remove the varying background due to the
continuum light from NGC~1275. The signal-to-noise ratio of
individual pixels is about 2 and we  investigate the
surface brightness distribution of pixels as follows:

The ds9 interface to Funtools was used to make histograms of the
number of pixels with different pixel values at various locations in
the nebula. Inspection of these histograms shows that there seems to
be a minimum surface brightness for the filaments, well above the
background noise level that occurs at a similar level throughout the
outer nebula.  A region with no filaments present shows an
approximately gaussian histogram centred around
zero with standard deviation $\sigma=3.27\times 10^{-3}$ (Fig. 4, left).

Data were extracted from a region of the Northern Filament, close to
that shown in Fig. 1.
The resulting histogram (Fig.~4, right)  can be well fit using the sum
of 6 gaussian functions. The first gaussian is constrained
to the parameters of the background, as measured above.  The
subsequent gaussians are constrained to have the same width as the
background (presumed to be counting statistics) but with centroids
corresponding to integer multiples of a minimum intensity value
measured somewhere in the nebula.

Although we caution that the fitted model is not unique, a ready
interpretation is that the surface brightness of individual filaments
is constant while the variation of surface brightness that we observe
in the image is due to the superposition of filaments (or
strands). The model assumes that all filaments are at the same
inclination to the observer, which may be reasonable for the Northern
Filament.  Our fit, shown as the solid line in Fig.~4 suggests a
decreasing trend in the sense for each increase in the number of
overlapping filaments there are approximately half as many pixels (or
half the area). We conclude that the data are consistent with a
roughly constant surface brightness for the threads, as expected by
our model.

\section{The reconnection diffusion process}

The interpenetration of hot plasma into filaments can occur through the
process of reconnection diffusion (Lazarian et al. 2010, Lazarian
2011). The process is easy to understand. In a laminar fluid magnetic
field lines maintain their identity and therefore particles entrained
on magnetic field lines stay there indefinitely, provided that Ohmic
resistivity is infinitely small. In a turbulent fluid, magnetic fields
are being driven perpendicular to their direction and cross each
other. A model of reconnection in Lazarian \& Vishniac (1999,
henceforth LV99) describes the process and shows how the reconnection
can be fast, i.e. independent of resistivity. As the LV99 predictions
were successfully tested in Kowal et al. (2009), it is sensible to
discuss the implications of the model for the filaments we deal with in the
paper.

Magnetic field lines from the hot region connect with magnetic fields
from the cold region as reconnection proceeds and therefore the hot
plasma intermingles with cold material. The excitation of emission in
this situation is mediated by reconnection which allows plasma
percolation into filaments, but its magnetic energy is not lost via
the process. The emission from filaments is only limited by the flow
of the hot gas around the filaments and the presence of turbulence
which induces fast reconnection and turbulent transport of magnetic
field and matter.

It is natural to assume that the turbulent velocity in filaments is
less than the Afven speed $v_{\rm A}$. (In the opposite limit one expects the
turbulence to be superAlfvenic and mix up magnetic fields on the eddy
turnover time scale, preventing the existence of long-lived
filaments.) The diffusion coefficient for turbulent transport of plasma
was estimated in Lazarian (2006):
\begin{equation}
\kappa_{\rm dyn}\approx Lv_{\rm L} M_{\rm A}^3
\end{equation}
where $L$ is the turbulence injection scale, $v_L$ is the turbulent
velocity at injection, $M_{\rm A}\equiv v_{\rm L}/v_{\rm A}$ is the
Alfvenic Mach number. The diffusion coefficient given by Eq. (1) is
applicable assuming that the size of the diffusion region $r>>L$. If the
radius of filaments is smaller than L, the diffusion coefficient
is modified to $\kappa_{\rm dyn}\approx r V_{\rm A} (r/LM_{\rm
  A}^3)^{1/3}$, provided that $r<LM_{\rm A}^3$.

The timescale for hot plasma to percolate into a filament of radius
$r$ is $\sim r^2/\kappa_{\rm dyn}$. In the case of $r\sim L$, the time
scale of plasma percolation is $\sim M_{\rm A}^{-3} r/v_{\rm
  L}$. These are the timescales over which the plasma and cold gas
mix. The flow of plasma into the cold gas can be estimated as
$nkTv_{\rm dt}$, where $v_{\rm dt}\approx \kappa_{\rm dyn}/r$.

We note that the linewidths of the filament emission are observed to
be moderately high (about $100\kmps$ for the outer filaments around
NGC\,1275, Hatch et al 2006). This is indicative of turbulence in the
filaments. The same observations show that the mean velocity field of
the filaments is fairly ordered, suggesting that the turbulence is
relatively small scale, of the order of the width of a filament.

The process that will compete with the reconnection diffusion is the
streaming of hot electrons along magnetic field lines. In any
turbulent magnetic field the diffusion of particles along magnetic field
lines is different from diffusion along laminar magnetic field
lines. The estimate in Lazarian (2006) provides for the diffusion of
electrons\footnote{This is probably an overestimate, as the electric field
  should prevent free streaming of electrons.}:
\begin{equation}
\kappa_{\rm e}\approx {1\over 3} v_{\rm e} \lambda M_{\rm A}^4
\end{equation}
where $v_{e}$ and $\lambda$ are the velocity and the mean free path of
an electron. The mean free path of electrons  $\lambda=300 T_3^2
n^{-1}_{-2}$~pc, where $T_3\equiv kT/(3 \keV)$ and $n_{-2}\equiv
n/(10^{-2} \pcm)$. The corresponding diffusion velocity of hot
electrons into filaments $v_{\rm de}\approx
r/\kappa_{\rm e}$ is lower than $v_{\rm dt}$ for the
numbers adopted\footnote{The reasons why the actual $\lambda$ may be
  substantially smaller than our estimate for the mean free path are
  provided in Lazarian et al. (2011; see also Brunetti \& Lazarian
  2011). This can only strengthen our
  conclusion that reconnection diffusion is the dominant source of
  injecting plasma into filaments.}

In summary, turbulent diffusion in which reconnection allows the hot
and cold plasma to intermingle can plausibly operate at the hot / cold
interfaces. The flow velocity is subAlfvenic in the cold gas or just a
few $\kmps$ (the Alfven velocity in the cold to warm gas is 1 to
$10\kmps$ if the magnetic field is close to being in pressure
equilibrium with the external hot thermal gas). At first sight this
does not match the velocity required by the arguments in Section 2,
where the flow velocity has to approach the sound speed $c_s$ in the
hot gas (300 -- $700\kmps$) to yield a filament flux $\sim p c_{\rm
  s}$. However, lack of velocity can be compensated for by increase
in surface area. We have argued that a filament is formed of many
smaller thread and strand-like components, which means that the total
area of the hot--cold interface within a filament is much larger than
that presented superficially to an outside observer. We require a
factor of 100 to 1000 increase which does not seem unreasonable. There
is of course an upper limit to how much flux is obtained from a
filament due to the energy flux in the external hot gas, which is of
the order of the saturated conduction flux of Cowie \& McKee,
introduced in Section 2.  The observations show that the filaments
around NGC\,1275 are within a factor of ten of that limit. The
observed optical linewidths are probably due to the spread in bulk 
velocities of the threads and strands within a filament.  

Further work is required to identify what keeps the filaments stable,
of similar size and presumably similar internal thready structure.

\section{Discussion and Conclusions}

We have shown that there is a fair coincidence between the
particle flux impinging on the cold filaments and the total radiation
flux emitted by them, as inferred from the observed H$\alpha$ line
emission. This provides good evidence that the secondary ``knock-on'' particles
which account for the observed line spectrum from the filaments
(Ferland et al 2009) originate from the interpenetration of the hot
intracluster gas with the cold filaments.

Turbulent diffusive reconnection provides a reasonable explanation for
overcoming the magnetic barrier which would otherwise keep the hot
and cold gasses separate.
Within the model of reconnection diffusion adopted in the paper
magnetic reconnection enables plasma initially entrained on different
field lines to come into contact. In other words, magnetic
reconnection in turbulent media removes magnetic barriers otherwise
existing between hot and cold gas. Alfvenic turbulence inducing
motions perpendicular to magnetic field and electrons streaming along
wandering magnetic field lines acts to induce efficient thermal conductivity.

Any fast magnetic reconnection process\footnote{Reconnection is fast
  when it does not depend on resistivity. In the Lazarian \& Vishniac
  (1999) model of magnetic reconnection, speed is determined by the
  turbulence injection scale and velocities. These dependences have
  been confirmed in numerical study by Kowal et
  al. (2009). Reconnection processes that depend on resistivity are
  mostly unimportant for astrophysical circumstances, as the magnetic
  field changes that they can induce are negligibly small.}
dissipates only a small fraction of energy through Ohmic
resistivity. The rest goes to straightening magnetic field lines
inducing motions of magnetized fluid. This can produce gas heating as
well as accelerate energetic particles either through second order
Fermi acceleration via the interaction of turbulence in the outflow
region with the particles (Larosa \& Shore 1998) or through the first
order Fermi acceleration as described in De Gouveia dal Pino \&
Lazarian (2005) (see  also Lazarian 2005 and Lazarian et
al. 2011). The efficiency of both heating and particle acceleration
depend on the amount of free energy that is being released via
reconnection. Both theory and numerical simulations (see Goldreich \&
Sridhar 1995, Lazarian \& Vishniac 1999, Cho \& Lazarian 2002) testify
that magnetic field lines getting more and more parallel to each other
with the decrease of the scale of the turbulent motions, thus making
less and less energy available to be dissipated via reconnection. In
other words, the process of reconnection that we appeal to in this
paper does not entail appreciable heating or particle acceleration. 

The accompanying X-ray emission is unexplained. The observed X-ray
spectrum of the Northern filament appears to be due to Fe-L and O
emission lines. The emission  could originate in gas at intermediate
temperatures (0.6-1.5~keV) between the hotter 4~keV gas in which the
cold ($T<5\times 10^4\K$) filaments are embedded, perhaps due to
mixing, or evaporation of some strands of cold gas.

Charge exchange is expected to operate however on the highly ionized
species interpenetrating the cold gas. We have examined the expected
emission from this, which would indeed be Fe-L and O lines emission,
but find that the predicted flux fails by a factor of 30, so is too
faint to account for the observations. This estimate is based on each
incoming ion having only one chance of charge exchange per ion
stage. If, as we have argued above, the cold threads are full of tiny
strands, it is plausible that an ion might partially recombine when
traversing a cold region and then undergo collisional ionization if it
then passes into hot gas again. The collisional ionization timescale
is only a few per cent of the crossing time of a filament. We
speculate that this may boost the charge exchange emission to the
observed level. (Note that charge exchange is not invoked to explain
  the optical, or infrared, emission from filaments.) 

Interpenetration of the cold gas by hot gas means that there is a flux
of $\sim 10^6$~particles~s$^{-1}$~cm$^{-2}$ into the filaments,
assuming the values for the N filament. The radiative cooling times of
the cold gas are much shorter than any other relevant times so 
  they imply a significant accretion of mass by the filaments.
Scaling with this value from the H$\alpha$ flux of the N filament to
the total H$\alpha$ luminosity observed from the filaments of
NGC\,1275 ($2.5\times 10^{42}\ergps$; Heckman et al 1989) indicates a
total mass accretion/cooling rate of $\dot M\sim 100\Msunpyr$.  ($\dot
M \approx70 L_{43} T_7^{-1}\Msunpyr,$ where the total luminosity of
the cold/cool gas (about 10--20 times the H$\alpha$ luminosity)
$L=10^{43}\ergps$ and the surrounding hot gas has a temperature of
$10^7\K$.) There is therefore a significant cooling flow proceeding
in the Perseus cluster, with the soft X-ray stage ($kT\sim 3\keV$ and
below) being carried out by the interpenetration/mixing of the cold
gas by the hot gas (as suggested by Fabian et al 2002 and Soker et al
2004). The ``missing soft X-ray luminosity'' of cool core clusters is
comparable to that which emerges at longer wavelengths through the
filaments (Fabian et al 2002). The mass cooling rate can adequately
balance the star formation rate over the outer filaments of
$20\Msunpyr$ (Canning et al 2010). The mass of the whole filamentary
nebulosity will double in about 1~Gyr. Note that we are not
  arguing that filaments only operate in the coolest parts of the hot
  gas but just take hot gas from where they happen to be. There may of
  course be a tendency for there to be more filaments where the hot
  gas is coolest.

We shall examine the situation for other emission line nebulae found
around brightest cluster galaxies in later work. The processes
outlined here should apply generally to cold gas embedded in hot or
energetic atmospheres such as found in nearby and distant radio
galaxies, Ly$\alpha$ nebulosities found at high redshift, the cores of
elliptical galaxies and possible LINERs.  They may also be relevant to
the H$\alpha$ emission from cold gas observed from galaxies being
stripped by the intracluster medium in other clusters (e.g. Sun et al
2008; Yagi et al 2010).
 
\section*{Acknowledgements}
We thank the referee, Ehud Behar, for helpful comments.
ACF thanks the Royal Society for support.  GJF acknowledges
support from NSF (0908877), NASA (07-ATFP07-0124 and 10-ATP10-0053)
and STScI (HST-AR-12125.01). AL acknowledges the support of the NSF
Center for Magnetic Self-Organization, NSF grant AST 0808118 and
NASA grant NNX09AH78G. This research  made use of
SAOImage DS9, developed by Smithsonian Astrophysical Observatory.

\bibliographystyle{mnras}

\begin{thebibliography}{}

\bibitem[]{} Baade W., Minkowski R., 1954, ApJ, 119, 215
\bibitem[]{} Begelman M.C.,  Fabian A.C., 1990, MNRAS, 244, 26
\bibitem[]{} B\"ohringerH.,  Fabian A.C., 1989, MNRAS, 237, 1147
\bibitem[]{} Branduardi-Raymont G., Bhardwaj A., Elsner R.F.,
  Rodriguez P., 2010, A\&A, 510, 73
\bibitem[]{} Bregman J.N., Fabian A.C., Miller E.D., Irwin J.A., 2006,
  ApJ, 642, 746 
\bibitem[]{} Brown G.V., et al 2009, J.Phys. Conf. Ser., 163, 012052
\bibitem[Brunetti 
\& Lazarian(2011)]{2011MNRAS.412..817B} Brunetti, G., \& Lazarian, A.\ 2011, MNRAS, 412, 817 
\bibitem[]{} Canning R.E.A., Fabian A.C., Johnstone R.M., Sanders
  J.S., Conselice C.J., Crawford C.S., Gallagher J.S., Zweibel E.,
  2010, MNRAS., 405, 115
\bibitem[Cho 
\& Lazarian(2002)]{2002PhRvL..88x5001C} Cho, J., \& Lazarian, A.\ 2002, Physical Review Letters, 88, 245001
\bibitem[]{} Crawford C.S., Fabian A.C., 1992, MNRAS, 259, 265
\bibitem[]{} Crawford C.S., Allen, S.W., Ebeling H., Edge A.C., Fabian
  A.C., 1999, MNRAS, 306, 857
\bibitem[]{} Conselice C.J., Gallagher J.S., Wyse R.F.G., 2001, ApJ,
  122, 2281
\bibitem[]{} Cowie L.L., McKee C.F., 1977, ApJ, 211, 135
\bibitem[de Gouveia dal Pino 
\& Lazarian(2005)]{2005AA...441..845D} de Gouveia dal Pino, E.~M., \&
Lazarian, A.\ 2005, A\&A, 441, 845 
\bibitem[]{} Dalgarno A., Yan M., Liu W., 1999, ApJS, 125, 237	
\bibitem[]{} Dennerl K., Englhauser J., Tr\"umper J., 1997, Sci, 277,
  1625
\bibitem[]{} Donahue M., Mack, J., Voit G.M., Sparks W., Elston R.,
  Maloney R., 2000, ApJ, 545, 670
\bibitem[]{} Edge A.C., 2001, MNRAS, 328, 762
\bibitem[]{} Edge A.C., Wilman R.J., Johnstone R.M., Crawford C.S.,
  Fabian A.C., Allen S.W., 2002, MNRAS, 337, 49 
\bibitem[]{} Fabian A.C., Allen S.W., Craford C.S., Johnstone R.M.,
  Morris R.G., Sanders J.S., Schmidt R.W., 2002, MNRAS, 332, L50
\bibitem[]{} Fabian A.C., Johnston R.M., Sanders J.S., Conselice C.J.,
  Crawford C.S., Gallagher J.S., Zweibel E., 2008, Nature, 454, 968
\bibitem[]{} Ferland G.J., Fabian A.C., Hatch N.A., Johnstone R.M.,
  Porter R.L., van Hoof P.A.M., Williams R.J.R., 2009, MNRAS, 392,
  1475 
\bibitem[]{} Ferland G.J., Fabian A.C., Hatch N.A., Johnstone R.M.,
  Porter R.L., van Hoof P.A.M., Williams R.J.R., 2008, MNRAS, 386, L72 
\bibitem[Goldreich 
\& Sridhar(1995)]{1995ApJ...438..763G} Goldreich, P., \& Sridhar, S.\ 1995, ApJ, 438, 763
\bibitem[]{} Hatch N., Crawford C.S.,  Fabian A.C.,  Johnstone R.M.,
  2005, MNRAS, 358, 765
\bibitem[]{} Hatch N., Crawford C.S.,  Johnstone R.M., Fabian A.C.,
  2006, MNRAS, 367, 433
\bibitem[]{} Heckman T.M., Baum S.A., van Breugel W.J.M., McCarthy P.,
  1989, ApJ, 338, 48
\bibitem[]{} Jaffe W., Bremer M.N., van der Werf P., 2001, MNRAS, 324, 443
\bibitem[]{} Johnstone R.M., Fabian A.C.,  Nulsen P.E.J., 1987, MNRAS,
  224, 75
\bibitem[]{} Johnstone R,N., Hatch N.A., Ferland G.J., Fabian A.C.,
  Crawford C.S., Wilman R.J., 2007, MNRAS, 382, 1246
\bibitem[]{} Knoll D.A., Brackbill J.U., PhPl, 9, 3775
\bibitem[]{} Kowal G., Lazarian A., Vishniac E.T., Otmianowska-Mazur,
  K., 2009, ApJ, 700, 63 
\bibitem[]{} Lallement R., 2004, A\&A, 422, 391
\bibitem[Larosa 
\& Shore(1998)]{1998ApJ...503..429L} Larosa, T.~N., \& Shore, S.~N.\ 1998, ApJ, 503, 429
\bibitem[]{} Lazarian A., Vishniac E.T., 1999, ApJ, 517, 700
\bibitem[]{} Lazarian A., 2006, ApJ, 645, L25
\bibitem[Lazarian et al.(2010)]{2010ASPC..429..113L} Lazarian, A., 
Santos-Lima, R., \& de Gouveia Dal Pino, E.\ 2010, 
Astronomical Society of the Pacific Conference Series, 429, 113
\bibitem[Lazarian(2005)]{2005AIPC..784...42L} Lazarian, A.\ 2005, Magnetic 
Fields in the Universe: From Laboratory and Stars to Primordial 
Structures., 784, 42 
\bibitem[Lazarian et 
al.(2011)]{2011P&SS...59..537L} Lazarian, A., Kowal, G., Vishniac, E., \& de Gouveia Dal Pino, E.\ 2011, PLANSS, 59, 537 
\bibitem[]{} Lynds R., 1970, ApJ, 159, L151
\bibitem[]{} Matsumoto Y., Hoshino M., 2006, JGRA, 1110, 213
\bibitem[]{} Nipoti C., Binney J., 2004, MNRAS, 349, 1509
\bibitem[]{} Salome P., Combes F., Revaz Y., Edge A.C., Hatch N.A.,
  Fabian A.C., Johnstone R.M., 2008, A\&A, 484, 317
\bibitem[]{} Sanders J.S., Fabian A.C., 2007,  MNRAS, 381, 1381
\bibitem[]{} Soker N., Blanton E.L., Sarazin C.L., 2004, 422, 445
\bibitem[]{} Sparks W.B., Pringle J.E., Carswell R., Voit M., Cracraft
  M., Martin R,G., 2009, ApJ, 704, L20
\bibitem[]{} Spitzer L., Tomasko M.G., 1968, ApJ, 152, 971
\bibitem[]{} Sun M., Donahue M., Roediger E., Nulsen P.E.J., Voit
  G.M., Sarazin C., Forman W., Jones C., 2010, ApJ, 708, 946
\bibitem[]{} Wilman R.J., Edge A.C., Johnstone R.M., Fabian A.C.,
  Allen S.W., Crawford C.S., 2002, MNRAS, 337. 63
\bibitem[]{} Yagi M., et al 2010, AJ, 140, 1814














\end{thebibliography}

\end{document}